\documentclass[]{aa}
\usepackage{graphicx}
\usepackage{natbib}
\usepackage{txfonts}

\begin{document} 
\title{Non-radial oscillations of the rapidly rotating Be star HD 163868}

\author{G.J. Savonije}

\institute{Astronomical Institute `Anton Pannekoek', University of
  Amsterdam, Kruislaan 403, 1098 SJ Amsterdam, The Netherlands}

\date{Received ; accepted }

\abstract 
{Oscillations in rotating stars with frequency $\bar{\sigma}$ of the same order or smaller than the rotation rate $\Omega$ cannot be described by a single spherical harmonic due to the effect of the Coriolis force. This is a serious complication which is usually treated by writing the eigenfunctions as a (truncated) sum of different spherical harmonic degrees for a given m-value, or by neglecting the $\theta$ part of the rotation vector (Traditional Approximation).}
{We aim for a more adequate treatment of the coupling with higher angular degrees for low frequency oscillations of rotating stars  by taking the Coriolis force fully into account, so that the coupling is included in the analysis (up to the grid resolution) and to compare the results with analyses based on the above mentioned approximations.}
{To this end a new, more efficient version of a 2D(r,$\theta$) implicit oscillation code was developed in which no a priori assumptions about the $\theta$ variation of the eigenfunctions is made, enabling a better treatment of the rotational truncation of g-modes near the stellar poles, and for which $\sim 150$ gridpoints in $\theta$ are feasible. We test the code by comparing the simulated oscillation spectrum  with that obtained by the MOST photometry for the  $\simeq$ 6 $M_\odot$ Be star  HD 163868.}
{We find both prograde and retrograde overstable modes (although more prograde than retrograde modes) and confirm the existence of low degree odd r-modes, destabilised by the $\kappa$-mechanism. The ultra-low frequency modes that could not be explained in a previous analysis are interpretated as (high degree) retrograde m=1 modes with $|\bar{\sigma}| \sim \Omega$ i.e. $\sigma \sim$ 0. A reasonably good fit to the observed oscillation spectrum is possible if we assume that only even modes are observed (no unstable r-modes visible). This requires a nearly equator-on view of the observed star, consistent with the measured high $v \, \sin i$ value of 250 km/s.}
{}
\keywords{Stars: emission-line, Be -- Stars: rotation, pulsation }

\maketitle

\section{Introduction}
In a previous paper \citep{Sav05} the pulsational stability of uniformly rotating main sequence stars in the mass range $3 -8\, M_\odot$ was studied by applying the traditional approximation (TA). Applying the TA renders the oscillation equations fully separable, greatly simplifying the solution, see e.g. \citet{unno89}. The earlier calculations indicated that the $\kappa$-mechanism in rapidly rotating stars can not only destabilise gravity (g) modes but, more surprising perhaps, also rotational r-modes. The TA analysis shows that the lowest degree odd (regarding the symmetry of the density and pressure perturbation with respect to the stellar equator) r-modes with low $m$-values have relatively large eigenvalues $\lambda$ for the angular part of the eigenfunctions. This indicates relatively large buoyancy effects for these r-modes, almost comparable to that of g-modes. The compressions and expansions in the driving zone (near $T \simeq 2 \times 10^5$ K) appeared indeed sufficiently strong for destabilization of some of these modes by the $\kappa$-mechanism.  This result was corroborated by the (TA) analysis of \cite{Town05}  and by the truncated spherical harmonics expansion method by \citet{Lee06}. Unfortunately the simplified TA analysis cannot take the coupling between different $l$-components through the Coriolis force adequately into account. 

\cite{SP97} developed an implicit two dimensional (2D) oscillation code, which solved the linear forced oscillation equations including the full Coriolis force (but neglecting the distorting centrifugal force), and applied it to the tidal evolution of massive binary systems. They found resonances with g-modes and r-modes and indications of interactions between the oscillations in the core and envelope when the forcing frequency $\bar{\sigma}$ (in corotating frame) was in the inertial regime ($|\bar{\sigma}| < 2 \Omega$), where $\Omega$ is the stellar rotation rate. 

Here we apply a similar implicit 2D code to the pulsational stability analysis of an evolved ($X_c=0.4$) 6 $M_\odot$ main sequence star rotating uniformly with $\Omega/\Omega_c=0.6$, where $\Omega_c$ is the surface break-up speed.  By applying a tidal force on the star with varying (complex) driving frequency one can search for resonances with free oscillation modes and determine their pulsational stability.  By varying the (sign of) the imaginary part  of the forcing frequency, and searching for a maximum response, we can determine the damping- or growth rate of the mode. This procedure is different from the usual stability analysis whereby one numerically integrates a (one dimensional) eigenvalue problem by iterating until one reaches a consistent solution. Explicit integration like this, whereby the boundary conditions are only fulfilled after a number of iterations, is likely to be impractical for the set of 2D partial differential equations describing stellar oscillations with the full Coriolis force. Therefore we apply a 2D code which numerically integrates the oscillation equations implicitly. The boundary conditions are applied along the rotation axis, along the stellar surface and back along the stellar equator. Starting at the stellar surface on the rotations axis, using the local boundary conditions together with the difference version of the oscillation equations, we start the elimination procedure that continues along all the points of the ($r$, $\theta$) grid towards the stellar centre. By finally applying the central boundary conditions we can then solve the unkown perturbations there and start a backsubsitution in reverse order towards the stellar surface.  The thus calculated result yields for each adopted forcing frequency (in the resonance search) a solution consistent with all boundary conditions, but it represents in general  not a free oscillation mode. However, when the frequency gets close to resonance, the forcing term becomes insignificant and the forced solution approaches that of the free oscillation mode.   
The implicit nature of this 2D integration causes the calculations to be memory and cpu-time expensive as the calculations involve many matrix inversions during the elimination procedure, the more so if one requires high grid resolution to allow adequate coupling with higher degree, short wavelength components. In order to improve the speed, and thus allow higher resolution of both the radial and angular part of the oscillatory response, a new efficient implicit 2D-code was developed based on a three-level difference scheme, see below.

 We compare our numerical results with the MOST observations of the Be star HD 163868 and the calculated oscillation spectrum  by \cite{Walker05} and \cite{Dziem07}.

\section{Non-adiabatic Oscillation Equations}
We introduce spherical coordinates ($r$, $\theta$, $\varphi$) centered at the stellar centre, whereby  $\theta=0$ corresponds to the stellar rotation axis, and solve the non-adiabatic linear oscillation equations with (the real part) of the  spherical harmonic forcing potential 
\begin{equation} \Phi^m_l=\Psi^m_l(r,\theta)\, \,\rm{e}^{\rm{i}(\sigma t- m \varphi)}= A \,\,r^l \, P^m_l(\theta) \,\, \rm{e}^{\rm{i}(\sigma t- m \varphi)} \label{forcing} \end{equation}
 where $A$ is a constant and $ P^m_l(\theta)$ the associated Legendre function. We define $m$ {\it to be always positive} and $\sigma$ the oscillation frequency in the inertial frame, see e.g. eqs. (1)-(4) in \cite{Sav05}. To keep the problem tractable we ignore the centrifugal distortion of the (rapidly) rotating star but include the full Coriolis force on the oscillations. Because the unperturbed state of the star is assumed spherically symmetric and we look for a stationary solution, the time and $\varphi$ part of the perturbed solution is separated as $\propto  \rm{e}^{\rm{i}(\sigma t- m \varphi)}$ and taken out of the equations below. For better efficiency we  pre-eliminate the $\varphi$ component of the displacement vector $\vec{\xi}$ together with the perturbed radiative flux components $F^\prime_r$ $F^\prime_\theta$ and $F^\prime_\varphi$. Hence only the four unknown perturbations $\xi_r$, $\xi_\theta$, $P^\prime/P$ and $\rho^\prime/\rho$ remain, where the symbols have their usual meaning and a prime denotes the Eulerian perturbation of a quantity.  We ignore perturbations of the convective energy flux. By defining $\bar{\sigma}=\sigma -m\, \Omega$, the oscillation frequency in the corotating frame (whereby $\bar{\sigma}<0$ corresponds to retrograde oscillations  and $\bar{\sigma}>0$ to prograde oscillations in stellar frame), the equation for the  perturbed radial motion of a mass element can be expressed as
\begin{eqnarray}
&& \left( \bar{\sigma}^2 -4 \, \Omega^2 \sin^2{\theta} \right) \, \xi_r -4 \, \Omega^2 \,\sin{\theta}\,\cos{\theta} \,\, \xi_\theta  + \frac{1}{\rho} \frac{\rm{d}P}{\rm{d}r} \, \left(\frac{\rho^\prime}{\rho}\right) + \nonumber \\
 && \left(\frac{2 m \Omega} {\bar{\sigma} r} \, \frac{P} {\rho} - \frac{1}{\rho}  \frac{\rm{d} P} {\rm{d} r} \right) \, \left(\frac{P^\prime}{P}\right) -\frac{P}{\rho} \frac{\partial}{\partial r} \left( \frac{P^\prime}{P}\right) =
- \frac{2 m \Omega}{\bar{\sigma} r} \, \Psi^m_l + \frac{\partial \Psi^m_l}{\partial r} \label{eq1}
\end{eqnarray}
and the equation of motion for the $\theta$ displacement
\begin{eqnarray}
&& \left(\bar{\sigma}^2 -4 \Omega^2 \, \cos^2{\theta} \right)\,\, \xi_\theta -\left(4 \Omega^2 \sin{\theta} \cos{\theta} \right) \,\, \xi_r  -\frac{P}{\rho r} \frac{\partial}{\partial \theta} \left(\frac{P^\prime}{P}\right) \nonumber \\
&& +\,  \frac{2 m \Omega}{\bar{\sigma} r} \frac{\cos{\theta}}{\sin{\theta}} \frac{P}{\rho} \, \left(\frac{P^\prime}{P}\right) = -\frac{2 m \Omega} {\bar{\sigma} r}  \frac{\cos{\theta}} {\sin{\theta}} \, \Psi^m_l + \frac{1}{r}\, \frac{\partial \Psi^m_l} {\partial \theta}, \label{eqn2}
\end{eqnarray}
while the perturbed equation for mass conservation follows as
\begin{eqnarray}
&& \frac{\rho^\prime}{\rho}+ \frac{1}{r^2 \rho} \frac{\partial}{\partial r} \left(r^2 \rho \, \xi_r\right) + \frac{1}{r \sin{\theta}} \frac{\partial}{\partial \theta} \left(\sin{\theta}\, \xi_\theta \right)+  \frac{2 m \Omega} {\bar{\sigma} r}  \, \xi_r 
+  \nonumber \\
&& \frac{2 m \Omega} {\bar{\sigma} r} \frac{\cos{\theta}}{\sin{\theta}} \,\, \xi_\theta - \left(\frac{m}{\bar{\sigma} r \sin{\theta}}\right)^2 \frac{P}{\rho}\, \left(\frac{P^\prime}{P}\right) = \left( \frac{m}{\bar{\sigma} r \sin{\theta}} \right)^2 \Psi^m_l.  \label{eqn3}
\end{eqnarray}

The perturbed energy equation with radiative diffusion can be written as (with $ \rm{i}$ the imaginary unit)
\begin{equation}
 \frac{P^\prime}{P} - \Gamma_1 \, \left(\frac{\rho^\prime}{\rho} \right) +\left( \frac{\rm{d} P}{\rm{d} r} - \Gamma_1 \frac{\rm{d}\ln \rho}{\rm{d} r} 
\right) \, \xi_r = \rm{i}\, \frac{\left(\Gamma_3-1\right)}{\bar{\sigma}\,\, P}\,\, \nabla \cdot \, F^\prime   \label{eqn4a}
\end{equation}
The divergence of the perturbed radiation flux is expressed as
\begin{eqnarray}
&& \frac{1}{F} \nabla \cdot F^\prime= - V ( 1 + \kappa_\rho)  \frac{\rho^\prime} {\rho}  - \frac{\partial}{\partial r}\left((1 + \kappa_\rho) \frac{\rho^\prime}{\rho} \right) +  V \left(4 - \kappa_T \right) \, \frac{T^\prime}{T}  \nonumber \\ 
&& + \frac{\partial}{\partial r} \left( (4-\kappa_T) \frac{T^\prime}{T} \right) +
\left( \Lambda V -\frac{ \frac{\rm{d}^2 \ln{T}} {\rm{d}^2 r}} {\left(\frac{\rm{d} \ln{T}} {\rm{d} r}\right)^2} \right) \frac{\partial}{\partial r} \left(\frac{T^\prime}{T}\right) + \Lambda \frac{\partial^2}{\partial^2 r} \left(\frac{T^\prime}{T}\right) + 
\nonumber \\
&& \frac{\Lambda} {r^2} \left[ -\left(\frac{m}{\sin{\theta}}\right)^2 \frac{T^\prime}{T}+ \frac{\cos{\theta}} {\sin{\theta}} \frac{\partial}{\partial \theta} \left( \frac{T^\prime}{T} \right) +  \frac{\partial^2} {\partial^2 \theta} \left(\frac{T^\prime}{T} \right) \right] 
\end{eqnarray}
where  $V=\frac{2}{r}+\frac{\rm{d} \ln{F}} {\rm{d}r} (\simeq 0$ in the envelope)  and $\kappa_\rho$, $\kappa_T$  denote the logarithmic derivatives of the opacity, while $\Lambda=(\frac{\rm{d} \ln{T}}{\rm{d} r})^{-1}$ the (negative) temperature scale height.
The unknown temperature perturbation $T^\prime/T$ (with its first and second order derivatives with respect to $r$ and $\theta$) is then eliminated through the perturbed equation of state
\begin{equation} \frac{P^\prime}{P}=\chi_T \,\, \frac{T^\prime}{T}+ \chi_\rho \,\, \frac{\rho^\prime}{\rho}-\chi_\mu \,\,\frac{\rm{d} \log \mu}{\rm{d} r}\,\, \xi_r 
\end{equation}
where the $\chi$ symbols denote as usual the partial logarithmic derivatives of the pressure with respect to temperature, density and mean molecular weight $\mu$. We have used the fact that diffusion may be neglected, so that the Lagrangian perturbation $\delta \mu =0$.
Using $P^\prime$ and $\rho^\prime$, instead of $T^\prime$ and $\rho^\prime$, gives a cleaner numerical solution in the regions with a (not perfectly smooth) $\mu$-gradient just outside the convective core. In the stellar interior the perturbations are almost adiabatic, so that the introduced complexities in the energy equation do not matter, whereas in the non-adiabatic surface regions the $\mu$-gradient vanishes.

\subsection{Differencing}
The remaining four partial differential equations are then approximated by four difference equations on a 2D grid of 3000 radial and 128 equidistant $\theta$ zones from $\theta=0$ to $\pi/2$. The four unknown perturbations  in the other hemisphere follow from symmetry (or anti-symmetry) about the stellar equator, depending on whether we force the star with odd or even $l-m$ values. The basic `difference atom' used for the 3-level differencing in  $(r$, $\theta)$ space is a cross with all four unknowns  $\xi_r$, $\xi_\theta$, $P^\prime/P$ and $\rho^\prime/\rho$ defined at the five levels corresponding to the centre and the four end points of the cross. This enables the treatment of second order derivatives in the energy equation and allows for the introduction of turbulent viscosity which, in convective regions, is added to the radial- and $\theta$- equation of motion. The viscosity is calculated with  simple mixing length theory, corrected with a reduction term (\cite{GK77}) when the convective timescale is larger than the oscillation period. The mixing length is taken as $\lambda_m=\rm{min}(2\, H_P,d_{min})$, where $d_{min}={\rm max}(|r-r_{cc}|,\, 0.25 H_P) \,\,$ with $H_P$ the pressure scale height and $r_{cc}$ the radius of the convective core boundary. The stellar model used for the unperturbed (spherically symmetric) star was obtained with a version of Eggleton's evolution code \citep{Pols95}.

\subsection{Stability Analysis with forcing}
We use a similar stellar model of a $6 M_\odot$ star with a core hydrogen abundance of $X_c=0.4$  with a radius of $4.34\, R_\odot$. We used the same OPAL opacities as \cite{Walker05}. With the newly developed (see above) 2D oscillation code working with a $(r, \theta$) grid it was determined for which complex frequencies $\bar{\sigma}$ forcing with a specific $l$ and $m$ component of the spherical harmonic forcing potential (\ref{forcing}) yields a maximum (resonant) response. This was accomplished by scanning through a range of real forcing frequencies and zooming in on apparent resonant frequencies and then optimising the stellar response (significantly)  by introducing an imaginary component in the forcing frequency.  Forcing at the (complex) resonance frequency $\bar{\sigma}_0$ excites (very nearly) the free oscillation mode in which we are interested, whereby $\mathcal{I}m( \bar{\sigma}_0)$ is a good estimate of the growth or damping rate of the free mode.  Forcing was performed with the lowest possible $l$ value for a given $m$ value and adopted equatorial (anti)symmetry. We consider only forcing with m=1 and m=2, so that the forcing has $l \le 3$. The number of nodes $n_\theta$ in the  range $0 < \theta < \pi$ of the harmonic forcing functions $\Psi^m_l$ is given by $n_\theta=l-m$. During the zoom-in procedure (close) to a resonance the oscillatory response often has a higher spherical degree than the forcing, e.g. $n_\theta$=4 for forcing with $n_\theta$=l-m=0. The converged oscillation mode (with complex frequency) may then have $n_\theta=2$. The excited oscillations for a given m-value  can in principle have any spherical degree consistent with the forcing symmetry (but is limited by the grid resolution). This complicates the searching procedure as the search may focus on a (stable) higher degree oscillation that occurs in the same frequency range and is excited through the  coupling caused by the Coriolis forces. Since it becomes only clear  {\it during} the zoom-in procedure on which kind of mode (spherical degree and stable or unstable) the search terminates, this cannot be prevented. The Coriolis coupling with higher degrees complicates the stability properties of the modes. It appears (see results section below) that the resonant overstable modes found are often of higher degree than the applied forcing. Some of the modes in an instability interval ($n_{min},n_{max}$) appear stable. This is caused by the stabilising coupling with higher degree/order modes.

\subsection{testcase: MOST observations}
\begin{figure}[t]
{\resizebox{0.5\textwidth} {!} {\rotatebox{-90}{\includegraphics{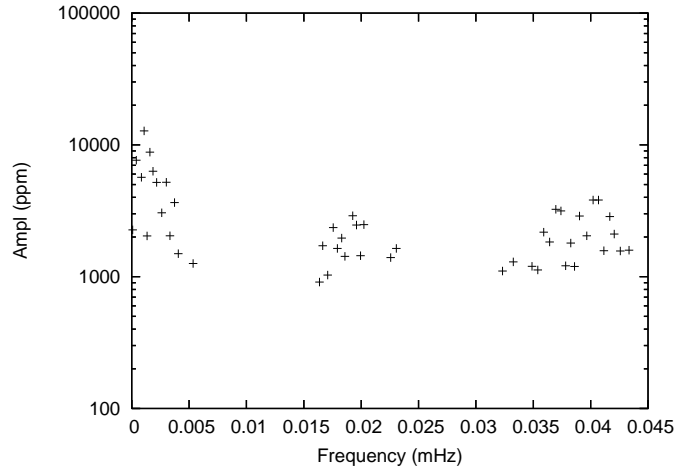}}}}
  \caption{The observed (MOST) amplitudes in mmag as a function of frequency in the inertial frame $f=\sigma/(2\, \pi)$ (Walker et al. 2005).}
  \label{most}
\end{figure}

\cite{Walker05} give an analysis of the photometric observations of the Be star HD 163868 with the Canadian MOST satellite in which three clusters of oscillation modes appear, see Fig.~\ref{most}. They  modelled stellar oscillations and searched for an  evolutionary model of the Be star which fits the oscillation spectrum: starting with a 6~$M_\odot$ ZAMS star with $X=0.7$ and $Z=0.02$ they arrived at a stellar rotation period of $\simeq 0.72$~d and evolution away from the ZAMS to a stellar radius of $R=4.3 R_\odot$. The opacity was obtained from the OPAL tables \citep{Igl96}. Their analysis (by H.S. and U.L, based on a truncated expansion in spherical harmonics) indicated that the oscillations with (inertial frame) frequencies  around 0.02 mHz and 0.04 mHz correspond to prograde, even g-modes with  $|m|=1$ and $|m|=2$, respectively.  They interpret the oscillations with frequencies near 0.005 mHz as due to (retrograde) r-modes with $|m|=1$. Apart from these latter modes, they only find pulsational instability for prograde modes. The very low frequency oscillations $f < 0.005$ mHz, where f=$\sigma /(2 \pi)$, could not be identified in their analysis. 

Very recently a new study \citep{Dziem07} based on the traditional approximation (TA) appeared which focusses on the oscillation spectrum of the same Be star. Below we discuss the results of both earlier analyses and compare them with our findings.

\section{Results} 

\subsection{Results for symmetric (even)  modes}
 We adopt a constant angular speed $\Omega=0.6 \, \Omega_c$ (with $\Omega_c=1.70 \times 10^{-4}$ sec$^{-1}$) corresponding to a rotation period of 0.71 d, like  \cite{Walker05} and, for some calculations, also $\Omega=0.7 \, \Omega_c$ . From now on we express all angular frequencies in units of $\Omega_c$.
\begin{figure}[htbp]
{\resizebox{0.5\textwidth} {!} {\rotatebox{-90}{\includegraphics{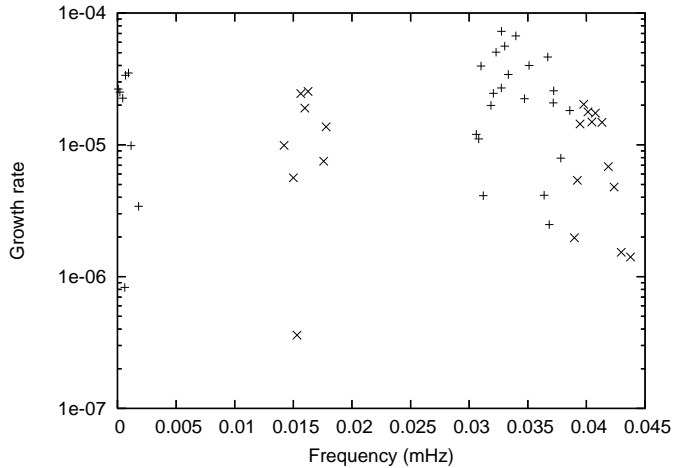}}}}
  \caption{The calculated growth rates $-\mathcal{I}m(\bar{\sigma})$ of overstable even oscillation modes versus frequency in the inertial frame for $\Omega=$~0.6, see Table~\ref{tab_even_06} for mode identification. Plusses: m=1; crosses: m=2}
  \label{even_06}
\end{figure}

\begin{table}[h]
\caption{Characteristics of the unstable even modes for $\Omega=0.6$ as shown in Fig.~\ref{even_06}.  The number of radial nodes (outside the convective core) in $\mathcal{I}m[\xi_r(r,\theta)]$ is given by $n_{r}$, while $n_{\theta}$ gives the number of nodes in the range $0 < \theta < \pi$ of $\mathcal{R}e[T^\prime(r,\theta)]$; $f_l$ and $f_u$ denote the lower and upper frequency of the instability interval  for the given $m$-value.}
\[ 
\begin{array}{ccccccc}  
\hline                                                                              
m &  n_\theta & n_r & \mathcal{R}e(\bar{\sigma})_u & \mathcal{R}e(\bar{\sigma})_l & f_l(mHz) & f_u(mHz)\\
\hline                                                                                             
\hline 
1 & 2   & 21;25:31  & -0.666 & -0.557 & -0.0018  & 0.0012\\
1 & 4   & 35        & 0.826  & -      &    -     &  0.0386 \\ 
1 & 2   & 18^{(1)}:40   & 0.797  & 0.531  &  0.0306 & 0.0364\\

\hline
2 & 2 & 21:28; 33:34 & -0.675 & -0.543 & 0.0135 & 0.0178\\
2 & 0 & 14:25   & 0.418  &  0.240 & 0.0390   & 0.0438 \\
2 & 2 & 25:31  & 0.732  &  0.618  & 0.0492 & 0.0523 \\
\hline            
\end{array}
\] 
$^{1}$: $n_\theta$=2 modes with $n_r$=19,20 are missing. Instead we find two unstable $n_\theta$=4 modes with $n_r$=40, 42. 
\label{tab_even_06}
\end{table}
\begin{figure}[]
{\resizebox{0.5\textwidth} {!} {\rotatebox{-90}{\includegraphics{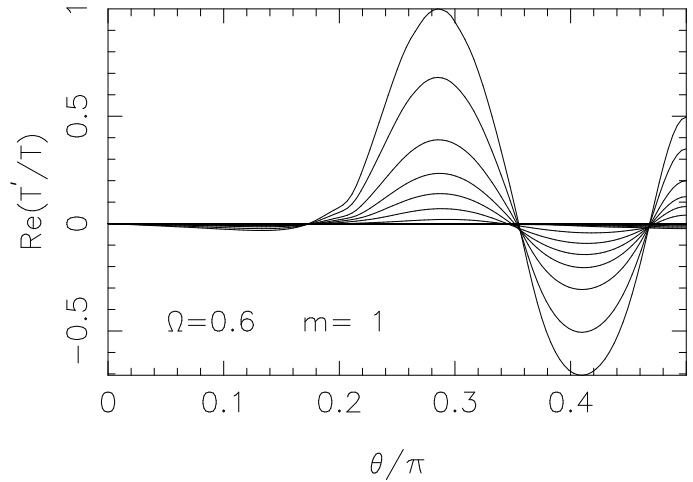}}}}
  \caption{Results for g$_{42}$, an unstable high degree  ($n_\theta=4$) m=1 prograde  g-mode with frequency f=0.0367 mHz in the observer's frame. Plotted is the $\theta$-variation of the relative temperature perturbation $T^\prime/T$  at 10 different radial shells in the stellar envelope. The largest amplitude curves correspond to the outermost layers. Adopted stellar rotation rate is $\Omega=0.6$.}
  \label{270313}
\end{figure}

\begin{figure}[t]
{\resizebox{0.5\textwidth} {!} {\rotatebox{-90}{\includegraphics{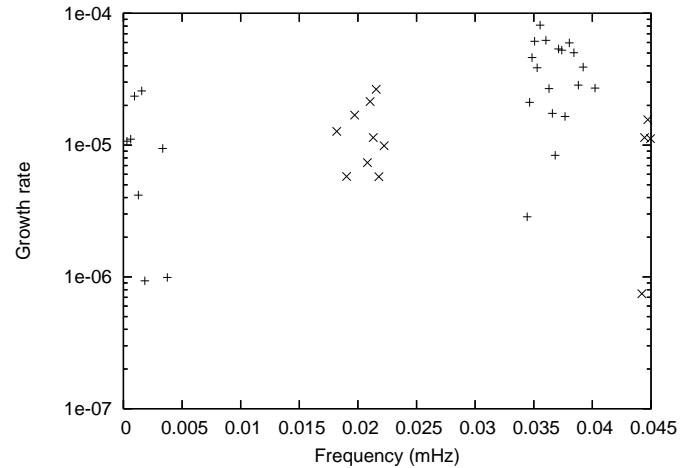}}}}
  \caption{The calculated growth rates $-\mathcal{I}m(\bar{\sigma})$ of overstable even oscillation modes versus frequency in the inertial frame for $\Omega=0.7$,  see Table~\ref{tab_even_07} for mode identification.  Plusses: m=1; crosses: m=2.}
  \label{even_07}
\end{figure}
The  calculated growth rates versus frequency in the inertial frame of overstable even oscillation modes are plotted in Fig.~\ref{even_06} for a stellar rotation frequency $\Omega$=0.6. Table~\ref{tab_even_06} lists the characteristics of the  unstable even modes found by the 2D code. It shows that almost all detected overstable even modes have higher angular degree than the applied forcing. The only exceptions are the lowest degree  ($n_\theta=0$)  prograde  m=2 modes with frequencies $\bar{\sigma}$ between 0.240-0.418.  We find a spectrum of much higher degree (than the forcing) resonances close to the unstable modes listed in tables~\ref{tab_even_06} and \ref{tab_even_07}. Usually these high degree/order modes are found to be stable, but not always. 
In  Fig.~\ref{270313} we show the temperature perturbation in the envelope for a high degree ($n_\theta$=4) unstable m=1 g-mode, listed in the footnote of table~\ref{tab_even_06}. The applied forcing has $l$=$m$=1, or $n_\theta$=0.
\begin{figure*}[]
{\resizebox{1.0\textwidth} {!} {\rotatebox{-90}{\includegraphics{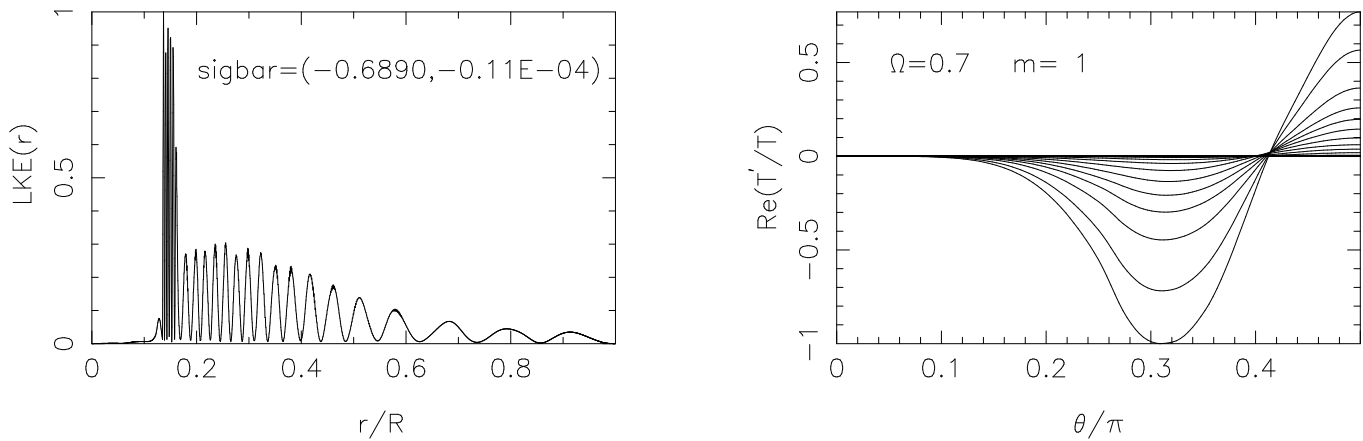}}}}
  \caption{Results for g$_{23}$, an even m=1 retrograde  g-mode with ultra low frequency f=0.0003 mHz in the observer's frame. In the panel on the left we plot the radial distribution of the local kinetic energy in the mode, while the righthand panel shows the $\theta$ variation (with $n_\theta=2$) of the relative temperature perturbation $T^\prime/T$  at 10 different radial shells in the stellar envelope. The largest amplitude curves correspond to the outermost layers. Adopted stellar rotation rate is $\Omega=0.7$.}
  \label{170244}
\end{figure*}
\begin{table}[]
\caption{Characteristics of the unstable even modes for $\Omega=0.7$ as shown in Fig.~\ref{even_07}. See Table~\ref{tab_even_06} for meaning of symbols.}
\[ 
\begin{array}{ccccccc}  
\hline                                                                              
m &  n_\theta & n_r & \mathcal{R}e(\bar{\sigma})_u & \mathcal{R}e(\bar{\sigma})_l & f_l(mHz) & f_u(mHz)\\
\hline                                                                                         \hline  
1 & 2 & 22:28 ; 35:36 & -0.703  & -0.562  & -0.0001 & 0.0037\\
1 & 2 & 23^{(1)}:40 &  0.749  & 0.573 & 0.0344 & 0.0392 \\
\hline
2 & 2 & 21^{(1)}:34 & -0.727  & -0.578 & 0.0182 & 0.0222 \\
2 & 0 & 15:26 & 0.387  &  0.234 & 0.0442   & 0.0484 \\
2 & 2 & 27^{(1)}:41^{(2)} & 0.736 &  0.585 & 0.0537 & 0.0578\\
\hline               
\end{array}
\]
$^1$: Some of the modes with $n_\theta=2$ seem to be missing. We do find (stable) higher degree modes in these intervals. \\
$^{2}$: the number of radial nodes is not well defined due to lack of resolution in the $\mu$-gradient layer
\label{tab_even_07}
\end{table}
The lowest degree ($n_\theta=0$) modes that were found to be overstable by Walker et al. (near $f \simeq 0.02$ mHz) were all stable according to our analysis. \cite{Dziem07} came to the same conclusion. 
Unlike Walker et al., we  do not find all  overstable g-modes to be {\it prograde}, although we find significantly more prograde than retrograde modes unstable. \cite{Dziem07}, on the other hand, found no significant difference  between prograde and retrograde modes. However, their analysis is based on the TA for which substantially more unstable modes are found than either by Walker et al. or by the present stability analysis.  Because the low frequency retrograde modes have an extra pair of angular nodes due to the presence of the branch of 'buoyant r-modes' the lowest angular order is absent, so that  there tend to be more prograde than retrograde unstable low frequency modes when this is not compensated by higher degree instability intervals. 
 The observed cluster of overstable modes near f=0.02 mHz are, in our interpretation, high radial order retrograde m=2 g-modes.

 We do find unstable low frequency modes with $f \le 0.005$ mHz that were observed by MOST and could not be explained by  Walker et al. These modes appear in our analysis as retrograde even m=1 g-modes with $n_\theta=2$, presumably of higher angular degree than searched for by Walker et al. Indeed, \cite{Dziem07} came to a similar interpretation of the very low frequency modes. An example of a ultra-low frequency m=1 mode (in observer's frame) with $\bar{\sigma} \simeq -\Omega$ can be seen in Fig.~\ref{170244}.

 In the left panel of Fig.~\ref{170244} we plot the local kinetic energy in consecutive  spherical shells of the grid  
\[ \rm{LKE}(r)=\bar{\sigma}^2\, \rho\,\int_\theta \int_\phi \left( (v^\prime_r)^2+(v^\prime_\theta)^2+(v^\prime_\phi)^2\right) \, r^2 \, \sin \theta\, \rm{d} \theta \, \rm{d}\phi \, \rm{d}r  \]
where $\rm{d}r$ is the radial gridsize. We observe that the kinetic energy of the mode peaks in the $\mu$-gradient layer just outside the convective core where the Brunt-V\"{a}is\"{a}l\"{a} frequency is large and the radial wavelength short. 
Due to rotational effects (Coriolis force) g-modes are confined to an equatorial band $\cos \theta \le |\bar{\sigma}|/(2 \,\Omega)$. This effect can be seen in the right panel of Fig.~\ref{170244}, where the amplitude in the (evanescent) polar regions of the star is almost zero. Although the $\theta$-distribution of the temperature perturbation of the retrograde m=1 modes (Fig.~\ref{170244}) indicates an angular order higher than the forcing, with $n_\theta=2$ (i.e. there are two nodes in the range $0 < \theta < \pi$) and the positive and negative parts cancel to some extent. Nevertheless, these higher degree modes may still be visible in the photometry even if we observe the star almost equator-on, this has to be studied in more detail, see \cite{Dziem07}.

It appears that several modes ($n_r$ values) inside a given instability interval are either missing or found stable, while their neighbours are again unstable, see tables~\ref{tab_even_06} and \ref{tab_even_07}. This phenomenon is absent when we apply the TA and use the same (now 1D) forcing method to analyse the pulsational stability. It is caused by the stabilising effect of Coriolis coupling with stable higher degree modes, as was also noted by \cite{Lee01}. Most of the damping occurs in the $\mu$-gradient layer just outside the convective core where the modes have short wavelengths and where the local kinetic energy of the modes is maximal. 

Comparing our results with the MOST observations, it can be seen in Fig.~\ref{even_06} that the unstable even retrograde m=1 modes all have inertial frame frequencies $f<~0.0012$~mHz, while the observed distribution in Fig~\ref{most} is spread over a larger frequency range. This can be improved by adopting a slightly higher rotation rate. We tried $\Omega$=0.7 for which,  in the inertial frame, the unstable m=1 modes are spread out to $\simeq 0.004$~mHz, more like what is observed, compare Figs.~\ref{most} and \ref{even_07}. However, the two clusters consisting of unstable retrograde m=2 modes (near f $\simeq$ 0.02 mHz) and prograde m=1 modes (near f$\simeq$ 0.04 mHz)  then shift to slightly too high frequencies. A rotation rate somewhere between 0.6 and 0.7 might give the best fit, although the observed mode clusters in Fig.~\ref{most} remain a bit more extended compared with the calculated spectrum.
Our interpretation requires that the MOST photometry detected only modes with {\it even} symmetry about the equatorial plane, i.e. HD 163868's  inclination angle should be near $\pi/2$. Cancellation effects would then explain that the overstable odd modes (in particular the r-modes) remain invisible in a photometric observation. A large inclination angle is consistent with the observed large value $v \, \sin i\simeq 250$ km/s for this star.

\subsection{Results for anti-symmetric (odd) modes}
\begin{figure*}[]
{\resizebox{1.0\textwidth} {!} {\rotatebox{-90}{\includegraphics{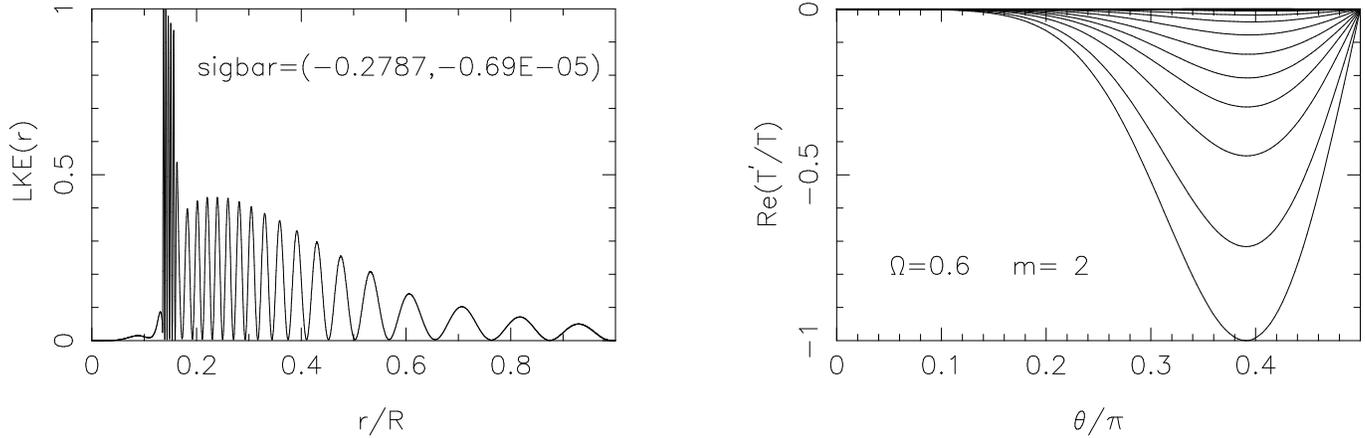}}}}
  \caption{Result for r$_{23}$ with f=0.0249 mHz, belonging to a series of unstable  m=2 buoyant r-modes. Adopted rotation rate is $\Omega$=0.6.}
  \label{040230}
\end{figure*}
\begin{figure}[htbp]
{\resizebox{0.5\textwidth} {!} {\rotatebox{-90}{\includegraphics{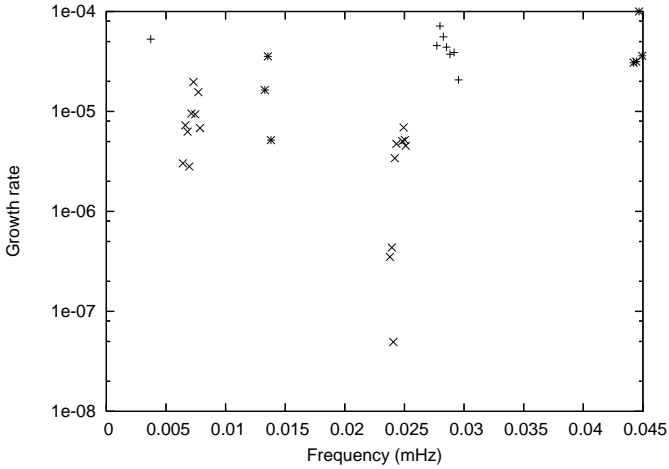}}}}
  \caption{The calculated growth rates $-\mathcal{I}m(\bar{\sigma})$ of overstable modes with odd symmetry versus frequency in the inertial frame for $\Omega=0.6$. See  Table~\ref{tab_odd_06} for mode identification. Crosses: buoyant r-modes with m=1 (left) and m=2 (right), plusses: g-modes with m=1 and asterisks: g-modes with m=2. }
  \label{odd_06}
\end{figure}
\begin{table}[h]
\caption{Characteristics of the unstable odd modes shown in Fig.~\ref{odd_06}, see Table~\ref{tab_even_06} for meaning of symbols. The retrograde modes with the lowest values of $|\mathcal{R}e(\bar{\sigma})|$ are the buoyant r-modes plotted in Fig.~\ref{odd_06}.}
\[ 
\begin{array}{ccccccc}  
\hline                                                                                             
m &  n_\theta & n_r & \mathcal{R}e(\bar{\sigma})_l & \mathcal{R}e(\bar{\sigma})_u & f_l(mHz) & f_u(mHz)\\
\hline                                                                                             
\hline                                                                                             
1 & 1 & 18:27 & -0.363 & -0.310 & 0.0064 & 0.0079\\
1 & 1 & 21:27 &  0.491  & 0.424 & 0.0295 & 0.0277\\
\hline
2 & 3 & 35:37^{(1)} & -0.709 & -0.690 & 0.0133 & 0.0138 \\
2 & 1 & 14:18;23:26 & -0.321 & -0.272 & 0.0238 & 0.0251\\
2 & 3 & \sim 35:37^{(1)} & 0.745 & 0.716 & 0.0526 & 0.0518 \\ 
2 & 1 & 17:33 & 0.679 & 0.434 & 0.0442 & 0.0524 \\
\hline                                                                                            
\end{array}
\] 
$^{1}$: the number of radial nodes is not well defined due to lack of resolution in the $\mu$-gradient layer
\label{tab_odd_06}
\end{table}
Fig.~\ref{odd_06} lists the unstable odd modes found in our calculation, see Table~\ref{tab_odd_06} for mode identification. For both m=1 and m=2 we found buoyant r-modes (or 'quasi g-modes') destabilised by the $\kappa$-mechanism, a result which corroborates earlier findings by \cite{Sav05} and \cite{Town05} with the traditional approximation and by \cite{Lee06}. These unstable r-modes are the retrograde modes with the smallest $|\bar{\sigma}|$ values in Table~\ref{tab_odd_06}. In Fig.~\ref{040230} we plot (right panel) the $\theta$-variation of a buoyant r-mode. The m=1 (odd) r-modes near $f~\sim ~0.005$ mHz were also found by  \cite{Walker05}.

Fig.~\ref{040230} shows that the high radial order buoyant r-mode exhibits a  $\theta$-variation similar to that of g-modes, although the ratio of $\xi_r/\xi_\theta$ is still significantly smaller than for g-modes. The 'quasi g-mode' in  Fig.~\ref{040230} shows considerable truncation near the pole, almost like a (normal) g mode.
For the odd modes we found, apart from the buoyant r-modes, only four  unstable retrograde modes (the three asterisks near f $\sim 0.013$ mHz and the plus near f$\sim$ 0.005 mHz in Fig.~\ref{odd_06}).

\section{Conclusions}
We interpret the whole oscillation spectrum derived from the MOST photometry \citep{Walker05} as being modes symmetric with respect to the stellar equator (even modes). It is assumed that the modes with odd equatorial symmetry are not observed because the Be star is seen almost equator-on, so that the light variations of the odd modes cancel to below the detection limit. The high value of $v \, \sin i \simeq 250$ km/s suggests indeed that the inclination angle $i$ is close to $\pi/2$. We identify the very low frequency (f$< 0.005$ mHz) oscillation modes in the observed spectrum (MOST) of HD 163868 as unstable retrograde m=1 oscillations and the intermediate cluster of modes around f=0.02 mHz as retrograde m=2 oscillations.   A similar conclusion was reached by \cite{Dziem07}. For $\Omega$=0.6 the highest frequency modes near f=0.04 mHz are identified as a combination of prograde m=1 and m=2 oscillations. For a higher rotation rate of $\Omega$=0.7, however, most of the prograde m=2 modes shift outside the frequency interval considered by \cite{Walker05}.  

The picture that emerges from the analyses by Walker et al (W05), Dziembowski et al (D07) and the present study  is not clear cut in that the detailed results differ in all three studies. W05 find a series of unstable prograde even m=1 modes that are all stable according to D07. We find the lowest angular order ($n_\theta$=0) m=1 also stable, but find unstable prograde m=1 modes with $n_\theta=2$. W07 find in general much more unstable modes than W05 or is found here. This is clearly linked to their use of the TA: in our test calculations with the TA  (and the same input model as for the 2D code) we also find more unstable modes, especially retrograde modes, than with the 2D code. This seems related to the fact that the TA does not take the coupling of different angular degrees by the Coriolis force properly into account. Both in W05's analysis and here significantly more prograde than retrograde modes are found unstable, contrary to the TA findings of D07. 

W05 uses Lee and Saio's method  in which the solution is approximated by a truncated set of spherical harmonics. However, the low frequency modes in rapidly rotating stars can deviate significantly from spherical harmonics, so that truncation can be problematic. In the 2D code used here no a priori assumptions about the $\theta$ behaviour of the solution (apart from boundary conditions) are made, i.e. no truncation occurs and the solution is limited only by the grid resolution. In this respect the present method seems therefore superior to the other two methods and does simulate the observed oscillation spectrum not in all details (e.g. the spread of the observed clusters of unstable modes is larger) but the overall result seems quite reasonable in view of the simplifications used (spherical, uniformly rotating star). The drawback of the implicit 2D method is its requirement of fast extended computer memory and patience of the user: a single frequency calculation (of which some 20 are required for a zoom-in on a resonance) takes about 40 minutes on a double Xeon processor node of the compute cluster LISA at SARA. 
 
A point of interest is the visibility of the modes as a function of inclination angle, especially those with the higher angular degrees, like the prograde m=1 modes with $n_\theta$=2.  See, e.g. \cite{Dziem07} for a discussion on  the visibility of modes.  This would need further exploration, together with a study of the effects of differential rotation.

\acknowledgement{We thank the referee (W. Dziembowski) for useful comments.}

\bibliographystyle{aa}
\bibliography{sav}

\end{document}